\documentclass{ifacconf}

\usepackage{natbib}
\usepackage{mathtools} 
\usepackage{amsmath,amssymb,amsfonts,subcaption,epsfig}
\usepackage{algorithm}
\usepackage{etoolbox}
\let\classAND\AND
\let\AND\relax
\usepackage{algorithmic}

\let\AND\classAND
\AtBeginEnvironment{algorithmic}{\let\AND\algoAND}

\floatname{algorithm}{Algorithm}

\usepackage{graphicx}
\usepackage{textcomp}
\usepackage{url}

\usepackage{pgfplots}
\usepgfplotslibrary{fillbetween} 
\usetikzlibrary{intersections}
\usetikzlibrary{calc, positioning}
\usepackage{epsfig}
\pdfoutput=1

\newsavebox{\ieeealgbox}

\def\comment#1{}
\newcommand{\R}{\mathbb{R}}
\newcommand{\N}{\mathbb{N}}

\def\bfm#1{\protect{\makebox{\boldmath $#1$}}}
\def\z {\bfm{z}}

\def\x {\bfm{x}}

\def\x{\bfm{x}}

\def\f{\bfm{f}}
\def\I{{\Bbb I \Bbb R}}
\DeclareMathOperator{\range}{range}

\def\f {\bfm{f}}

\newtheorem{remark}{Remark}

\newtheorem{definition}{Definition}
\newtheorem{corollary}{Corollary}

\def\bfm#1{\protect{\makebox{\boldmath $#1$}}}

\newtheorem{example}{Example}

\begin{document}

\begin{frontmatter}


\title{Tractable higher-order under-approximating AE extensions for non-linear systems}


\author[First]{Eric Goubault}
\author[First]{Sylvie Putot}

\address[First]{LIX, CNRS, Ecole Polytechnique, Institut Polytechnique de Paris, France (email: author@lix.polytechnique.fr)}


\begin{abstract}
We consider the problem of under and over-approximating the image of general vector-valued functions over bounded sets, and apply the proposed solution to the estimation of reachable sets of uncertain non-linear discrete-time dynamical systems. Such a combination of under and over-approximations is  very valuable for the verification of properties of embedded and cyber-physical controlled systems. Over-approximations prove properties correct, while under-approximations can be used for falsification. Coupled, they provide a measure of the conservatism of the analysis. 
This work introduces a general framework relying on computations of robust ranges of  vector-valued functions. This framework allows us to extend for under-approximation many precision refinements that are classically used for over-approximations, such as affine approximations, Taylor models, quadrature formulae and preconditioning methods. We end by evaluating the efficiency and precision of our approach, focusing on the application to the analysis of discrete-time  dynamical systems with inputs and disturbances, on different examples from the literature.
\end{abstract}

\end{frontmatter}

\section{Introduction}

Guaranteed state estimation and reachability analysis are central to many problems in control, such as robust and optimal control of dynamical systems, set invariance, safety verification, or control synthesis. 
This ultimately relies on computing ranges of functions over a domain, that we have to approximate since this is an intractable problem. 

Much of the existing work focuses on over-approxima\-tions of images of functions, or of reachable sets, generally based on convex set representations (in particular intervals, ellipsoids, polyhedra). We are interested here in the much less studied problem of computing under-approximations, that is, sets of states guaranteed to be reached. 
Combining over and under approximations is fundamental for the validation of control systems.
When the over-approximation is not sufficient to prove a property, an under-approximation is helpful to state the quality of the over-approximation. Additionally, when  an under-approximation of the reachable set intersects the set of error states, it  provides a proof of falsification of the property.

For general controlled systems, the reachability properties will depend on the initial conditions of the system, but also on the sensitivity of the system to some control inputs and external disturbances, as reflected by the notions of minimal and maximal reachability~\cite{Mitchell07}. We generalize these notions here to robust reachability, when both control inputs and adversarial disturbances are present. The robust image will be the intersection, for all possible disturbances, of the images of a function or reachable sets of a system.

\subsubsection*{Contents and contributions}


The computation of the reachable set of a dynamical system can be reduced to a series of images of sets by some vector-valued function. In previous work~\cite{lcss2020}, we  introduced mean-value extensions allowing us to compute under approximations (also called inner-approximations) of such images in a very efficient way. 
In this article, we generalize this approach to higher-order extensions, and develop quadrature formulas for more precise under-approximations. We also address many questions that are to be solved for an accurate and efficient implementation: 
\begin{itemize}
\item Section \ref{section1} recaps the necessary background from previous work. Section \ref{subsec:genAE} generalizes the mean-value extension of ~\cite{lcss2020} for the under and over-approximation of the robust range of sufficiently smooth real-valued functions $f: \R^m \rightarrow \R$;  this generalization allows us to propose  higher-order Taylor extensions for under-approximating robust ranges of functions. These new extensions, just as the mean-value extensions, is the basis for under-approximations of elementary vector-valued functions from $\R^m$ to $\R^n$, as detailed in Section \ref{sec:vectorfunctions}, which is instrumental in the reachability analysis of dicrete-time nonlinear systems proposed later;
\item   Section \ref{sec:quadra} proposes a novel approach to subdivisions for mean-value and Taylor extensions, based on the idea of quadratures (in numerical calculations of integrals): we show that this  improves precision of the computation of under and over-approximations, while still scaling with the dimension of the system;
\item Section~\ref{sec:discrete_systems} applies this work to the  approximation of robust reachable sets of discrete-time dynamical systems; we present   results on representative  systems from the literature, demonstrating the tractability and precision of our approach.
\end{itemize}  



\subsubsection*{Related work}
Our approach is related to and partially relies on work on modal intervals and mean-value extensions, which applications include the computation of under-approximations of function images~\cite{gold2,gold3}. 
It is also related to over-approximations of nonlinear functions and dynamical systems, on which we rely to compute under-approximations. Many methods for over-approximating reachable sets for non-linear systems have been developed, among which Taylor methods~\cite{Berz} or polytopes-based methods \cite{Girard2,Dreossi_2016}. 

There exist less methods for the harder problem of under-approximating images of functions or sets of reachable states. Some approaches have been proposed for linear discrete-time systems~\cite{ellipsoids00,girard2006,RAKOVIC200815327}.
Interval-based methods, relying on space discretization, have been used for under-approximating the image of nonlinear functions~\cite{Jaulin}. 
They were also used to over and under approximate solutions of differential systems with uncertain initial conditions~\cite{LEMEZO201870}. 
Tight approximations for reachable sets of nonlinear continuous systems can be found via expensive Eulerian methods: the zero sub-level set of the Lipschitz viscosity solution to a Hamilton-Jacobi (HJB) partial differential equation gives the (backward) reachable set~\cite{efficient-HJB-tomlin}. Other approaches, using SoS methods and LMI relaxations have been proposed for inner approximations, see e.g. \cite{Henrion}. In \cite{Xue4}, under-approximations for polynomial systems are obtained by solving semi-definite programs.
Taylor models are used on the inverse flow map to derive under-approximations~\cite{Sriram1}, but using topological conditions that are checked with interval constraints solving, which have difficulties to scale up with dimension. 
In \cite{Xue}, the computation of the under-approximated reachable set is based on a costly analysis of the boundary of the reachable sets and polytopic approximations. 
In \cite{Althoff20}, some non-convex under-approximations are computed with polynomial zonotopes, relying on a computation of the outer-approximation of the reachable set, of an enclosure of the boundary of the reachable set, and a reduction of the outer-approximation until it is fully included in the region delimited by the boundary.

\subsubsection*{Notations}

For a continuously differentiable vector-valued function $f: \R^m \rightarrow \R^n$, we note $f_i$ its $i$-th component and $\nabla f = (\nabla_j f_i)_{ij} = (\frac{\partial f_i}{\partial x_j})_{1\leq i \leq n, 1 \leq j \leq m}$ its Jacobian matrix. We note $\langle x, y \rangle$ the scalar product of vectors $x$ and $y$, and $\lvert x \rvert$ the absolute value extended componentwise. 

Intervals are used in many situations to rigorously compute with interval domains instead of reals, usually leading to over-approximations of function ranges over boxes. Set valued quantities, whether scalar or vector-valued, will be noted with bold letters, e.g $\x$. We denote $\I= \{\x=[\underline{x},\overline{x}] , \: \underline{x} \in \R,  \: \overline{x} \in \R \}$ the set of intervals with real-valued bounds. If $\overline{x} < \underline{x}$, the interval represents the empty set.  For a (possibly vector-valued) interval $\x \in \I^m$, we note $c({\x}) = (\underline{x}+\overline{x})/2$ its  center  and  $r({\x}) = (\overline{x}-\underline{x})/2$ its radius.
The operators over real numbers are lifted in intervals using the same notation.

\section{Background: AE extensions for computing function images}
\label{section1}

We recall in this section the notations and results of \cite{lcss2020}: we state
 mean-value over and under-approximating extensions for scalar and vector-valued functions. 

An \emph{over-approximating extension}, also  called \emph{outer-appro\-ximating extension}, of a function $f: \R^m \rightarrow \R^n$ is a function
$\f: \mathcal{P} (\R^m) \rightarrow \mathcal{P}(\R^n)$, such that for all $\x$ in $\mathcal{P} (\R^m)$, 
$ \mbox{range}(f,\x)=\{f(x), x \in \x\} \subseteq \f(\x)$. Dually, under-approximations determine a set of values proved to belong to the range of the function over some input set.  An \emph{under-approximating extension}, also called \emph{inner-approximating extension}, of $f$ is a function $\f: \mathcal{P} (\R^m) \rightarrow \mathcal{P}(\R^n)$, such that for all $\x$ in $\mathcal{P} (\R^m)$, 
$ \f(\x) \subseteq  \mbox{range}(f,\x)$. Under- and over-approximations can be interpreted as quantified propositions:   
$\range(f,\x) \subseteq \z$ can be written 
$ \forall x \in \x, \, \exists z \in \z, \, f(x)=z,$
while $\z \, \subseteq \range(f,\x)$ can  be written 
$ \forall z \in \, \z, \, \exists x \in \x, \, f(x)=z. $
Both these propositions are what we will call \emph{AE propositions}, for quantified propositions where universal quantifiers (A) precede existential quantifiers (E).

\subsection{Mean-value AE extensions for scalar-valued functions} 
\label{sec:meanvalueAE1D}

We  consider a function $f: \R^m \rightarrow \R$.
The natural interval extension consists in replacing real operations by their interval counterparts in the expression of the function. 
A generally more accurate extension relies on a linearization by the mean-value theorem. 

\subsubsection{Mean-value AE extensions}
Suppose $f$ is differentiable over the box $\x$. 
The mean-value theorem implies that  $$\forall  x^0 \in \x, \,  \forall x \in \x,
\exists \xi \in \x, \, f(x) = f(x^0) + \langle \nabla f(\xi), x-x^0 \rangle.$$
If we can bound the range of the gradient of $f$ over $\x$, by $ \bfm{ \nabla f}(\x)$, then we can 
derive an interval enclosure, called the mean-value extension.
Let us  choose $x^0$ to be the center $c({\x})$ of $\x$ and recall we note $r({\x}) = (\overline{x}-\underline{x})/2$ its  radius.
\begin{thm}[Thm. 1, \cite{lcss2020}]
\label{lemma1}
Let $f$  be a continuously differentiable function from $\R^m$ to $\R$ and  $\x \in \I^m$. Let 
$\f^0=[\underline{f^0},\overline{f^0}]$ include $ f(c({\x}))$ and $\bfm{\nabla}$ a vector of intervals $\bfm{\nabla_i} = [\underline{\nabla}_i,\overline{\nabla}_i]$ for $i \in \{1,\ldots,m\}$  such that 
 $\left\{ \left\lvert \nabla_i f (c(\x_1),\ldots,c(\x_{i-1}),x_i,\ldots,x_m) \right\rvert, x \in \x \right\} \subseteq \bfm{\nabla}_i.$ 
We have the over- and under-approximating extensions
 \begin{align}
 \mbox{range}(f,\x) \subseteq  [\underline{f^0},\overline{f^0}] +  \langle \overline{\nabla} , r({\x}) \rangle [-1,1]  \label{MV1} \\
[\overline{f^0} -  \langle \underline{\nabla} , r(\x) \rangle , \underline{f^0} + \langle \underline{\nabla} , r(\x) \rangle ] \subseteq \mbox{range}(f,\x)   \label{MV2}
\end{align}
\end{thm}


\begin{example}
\label{ex1}
Let us consider the range of $f$ defined by $f(x)=x^2-x$ over $\x=[2,3]$. We can compute $f(2.5)=3.75$ 
and $\lvert \nabla f([2,3]) \rvert \subseteq  [3,5]$. 
Then (\ref{MV1}) and (\ref{MV2}) yield
$ 3.75 + 1.5 [-1,1] \subseteq \mbox{range}(f, [2,3]) \subseteq 3.75 + 2.5 [-1,1],$
from which we deduce $[2.25,5.25] \subseteq \mbox{range}(f, [2,3]) \subseteq [1.25,6.25]$.
\end{example}

We  refer to  extensions (\ref{MV1}) and (\ref{MV2})  as \emph{AE extensions}, as they can be interpreted as \emph{AE propositions}. 
Note that the wider, lesser quality, are the over-approximations of $f$ and its derivatives, the tighter, less quality, are the under-approximations. The under-approximation can even become empty if the width $\overline{f_0}-\underline{f_0}$ of the approximation of $ f(c({\x}))$ exceeds $2 \langle \overline{\nabla} , r({\x}) \rangle $: in this case the lower bound of the resulting interval is larger than the upper bound, which by convention we identify with the empty interval.
Note also that when $ 0 \in \bfm{\nabla_i \f}$, 
then $\underline{\nabla_i} =0$ and if this is the case for all $i$, the under-approximation is empty or reduced to a point. A special attention to the practical evaluation of these extensions over the region $\x$ of interest is thus crucial, this is the object of Section~\ref{sec:quadra}.

\subsubsection{Mean-value AE extensions of the robust range}
\label{subsubsec:robustmeanvalue}
Mean-value AE extensions can be generalized to compute ranges that are robust to disturbances, identified as some input components.
Let us partition the indices of the input space in two subsets $I_{\mathcal{A}}$ and $I_{\mathcal{E}}$, where $I_{\mathcal{A}}$ defines the indices of the inputs that correspond to disturbances, and $I_{\mathcal{E}}$ the remaining dimensions. We decompose the input box $\x$ accordingly by $\x = \x_{\mathcal{A}} \times \x_{\mathcal{E}}$. 
We define the robust range of function $f$ on $\x$, robust on $\x_{\mathcal{E}}$ with respect to disturbances $\x_{\mathcal{A}}$, as
$ \mbox{range}(f,\x,I_\mathcal{A},I_\mathcal{E}) =  \{ z \, | \,  \forall w \in \x_{\mathcal{A}}, \, \exists u \in \x_{\mathcal{E}}, \, z = f(w,u) \}  $. Intuitively, $u$ will be control components, $w$ disturbances to which the output range should be robust.

\begin{thm}[Thm. 2, \cite{lcss2020}]
\label{prop:robustAE}
Let $f$ be continuously differentiable function from $\R^m$ to $\R$ and $\x = \x_{\mathcal{A}} \times \x_{\mathcal{E}} \in \I^m$. 
Let $\f^0$, $\bfm{\nabla}_{w}$ and  $\bfm{\nabla}_{u}$ be vectors of intervals such that $ f(c(\x)) \subseteq \f^0  $, $\{ \left| \nabla_w f (w,c(\x_{\mathcal{E}})) \right| \, , \, w \in \x_{\mathcal{A}} \} \subseteq  \bfm{\nabla}_{w}$ and $\{ \left| \nabla_u f (w,u) \right| \, , w \in \x_{\mathcal{A}} , \, u \in \x_{\mathcal{E}} \} \subseteq  \bfm{\nabla}_{u}$.
We have:
\begin{multline}
\label{MV1c}
 \mbox{range}(f,\x,I_\mathcal{A},I_\mathcal{E}) \subseteq  [\underline{f^0} - \langle \overline{\nabla}_u ,  r({\x_\mathcal{E}}) \rangle + \langle \underline{\nabla}_w ,  r({\x_\mathcal{A}}) \rangle , \\ \overline{f^0} + \langle \overline{\nabla}_u , r({\x_\mathcal{E}}) \rangle - \langle \underline{\nabla}_w , r({\x_\mathcal{A}}) \rangle ] 
 \end{multline}
 \vspace*{-0.7cm}
 \begin{multline}
\label{MV2c}
[ \overline{f^0}  - \langle \underline{\nabla}_u , r({\x_\mathcal{E}}) \rangle + \langle \overline{\nabla}_w , r({\x_\mathcal{A}}) \rangle ,  \underline{f^0} + \\ \langle \underline{\nabla}_u , r({\x_\mathcal{E}}) \rangle - \langle \overline{\nabla}_w ,  r({\x_\mathcal{A}}) \rangle ]  \subseteq \mbox{range}(f,\x,I_\mathcal{A},I_\mathcal{E})  
\end{multline}
\end{thm}



We refer to Example 2 of \cite{lcss2020} for a sample computation.

\subsection{AE extensions for vector-valued functions}
\label{sec:vectorfunctions}
Following \cite{lcss2020}, we now detail how full n-dimensional boxes can be included in the image of vector-valued functions $f : \R^m \rightarrow \R^n$, for $m \geq n$, using AE extensions of robust ranges. Theorem~\ref{prop_joint} and Definition~\ref{def1} will be instrumental in Algorithm~\ref{alg:discrete1} for discrete-time reachability of Section~\ref{sec:discrete_systems}.

The mean-value extensions of Theorem~\ref{lemma1} or the generalization of Theorem \ref{thm:robustapprox} 
give us under and over-approximations of projections of the image of the function. The Cartesian product of the over-approximations of each component provides an over-approximation of a vector-valued function $f : \R^m \rightarrow \R^n$. This is however not the case for under-approximation. Suppose for example that we have 
$\forall z_1 \in \z_1, \exists x_1 \in \x_1, \, \exists x_2 \in \x_2,  \: z_1 = f_1(x)$ and
$\forall z_2 \in \z_2, \exists x_1 \in \x_1, \, \exists x_2 \in \x_2,  \: z_2 = f_2(x) $.
We cannot deduce directly that for all $\forall z_1 \in \z_1$ and $\forall z_2 \in \z_2$ there exists $x_1$ and $x_2$ such that $z=f(x)$. 

Suppose now that we have: 
$\forall z_1 \in \z_1, \forall x_1 \in \x_1, \, \exists x_2 \in \x_2,  \: z_1 = f_1(x)$ and $\forall z_2 \in \z_2, \forall x_2 \in \x_2, \, \exists x_1 \in \x_1,  \: z_2 = f_2(x)$
\noindent with continuous selections $x_2$ and $x_1$. 
Then there exists
functions $g_2(z_1,x_1): \z_1 \times \x_1 \rightarrow \x_2$
and $g_1(z_2,x_2): \z_2 \times \x_2 \rightarrow \x_1$ that are continuous in $x_1$ (resp. $x_2$), 
and such that $\forall (z_1,z_2) \in \z$,  $\forall (x_1,x_2) \in \x$,  
$z_1 = f_1(x_1,g_2(z_1,x_1))$ and $z_2 = f_2(g_1(z_2,x_2),x_2) $.
Using the Brouwer fixed point theorem on the continuous map $g~: (x_1, x_2) \rightarrow (g_1(z_2,x_2),g_2(z_1,x_1))$ on the compact set $\x_1 \times \x_2$, then $\forall (z_1,z_2) \in \z$,  $\exists (x^z_1,x^z_2) \in \x $ such that $(z_1,z_2) = f(x^z_1,x^z_2)$. 

This result can be generalized to functions $f~: \R^m \rightarrow \R^n$ for any $n$, as stated in Theorem~\ref{prop_joint}. 


\begin{thm}[Theorem~3 in \cite{lcss2020}]
\label{prop_joint}
Let $f: \R^m \rightarrow \R^n$  be an elementary function and $\pi : [1 \ldots m] \rightarrow [1 \ldots n]$
Let us note, for all  $i \in [1 \ldots n]$, 
$J_E^{(z_i)} = \{ j \in  [1 \ldots m], \; \pi(j) = i \} $ and $J_A^{(z_i)} = \{ j \in  [1 \ldots m] \} \setminus J_E^{(z_i)}$.
Consider the $n$ AE-extensions $i \in [1 \ldots n]$ built from Theorems \ref{prop:robustAE}, \ref{thm:generalAE} or \ref{thm:robustapprox} and such that 
\begin{equation}
\forall z_i \in \z_i, \:  (\forall x_j \in \x_j)_{j \in J_A^{(z_i)}}, \:  (\exists x_j \in \x_j)_{j \in J_E^{(z_i)}}, \: z_i = f_i(x) 
\label{AEi}
\end{equation}
Then  $\z = \z_1 \times \z_2 \times \ldots \times \z_n$, if non-empty, is an under-approximation of the image of $f$:
$ \forall z \in \z,  \,  \exists x \in \x, \: z = f(x).$
\end{thm}


Theorem~\ref{prop_joint} gives us directly a computation of an under-approximation of $\mbox{range}(f,\x)$ for  $f: \R^m \rightarrow \R^n$. It can also be used to compute an under-approximation of the robust range $\mbox{range}(f,\x,I_\mathcal{A},I_\mathcal{E})$. For this, we need to choose $\pi : [1 \ldots m] \rightarrow ([1 \ldots n] \setminus I_\mathcal{A})$, which corresponds to the fact that the disturbance part of the input components will always be quantified universally. We define below the result of this process, which will be later used in reachability algorithms for discrete-time dynamical systems. 
\begin{definition}
\label{def1}
Let $f: \R^m \rightarrow \R^n$ and $\pi : [1 \ldots m] \rightarrow ([1 \ldots n] \setminus I_\mathcal{A})$. We define ${\cal I}(f,\x,I_\mathcal{A},I_\mathcal{E},\pi)$ an under-approximation of  $\mbox{range}(f,\x,I_\mathcal{A},I_\mathcal{E})$ obtained using Theorem~\ref{prop_joint} with function $\pi$, in which the under-approximation of each component is obtained using Theorem~\ref{prop:robustAE} (or Corollary~\ref{corollary1}).
We define ${\cal O}(f,\x,I_\mathcal{A},I_\mathcal{E},\pi)$ the over-approximation of the robust range obtained using Theorem~\ref{prop:robustAE} component-wise.
\end{definition}

\section{Generalization to new AE extensions}

\label{subsec:genAE}

We now introduce new  robust AE extensions for a function $f  :  \R^m \rightarrow \R$, which are no longer necessarily based on the mean-value theorem. 
\begin{thm}
\label{thm:generalAE}
Suppose we have an approximation function $g$ for $f$, which is an elementary\footnote{Elementary functions are compositions of +, -, $\times$, /, sine, cosine, log, exp functions in particular.} function in the sense of \cite{gold2}, satisfying
$\forall w \in {\bf x}_{\cal A}, \ \forall u \in {\bf x}_{\cal E}, \ \exists \xi \in {\bf x}, \ 
f(w,u)=g(w,u,\xi)$.
\noindent Then any under-approximation (resp. over-approximation) of the robust range of $g$ with respect to $x_{\cal A}$ and $\xi$, ${\cal I}_g \subseteq \mbox{range}(g, {\bf x}\times {\bf x}, I_{\cal A}\cup \{m+1,\ldots,2m\}, I_{\cal E})$ is an under-approximation (resp. over-approximation) of the robust range of $f$ with respect to $x_{\cal A}$, i.e. ${\cal I}_g \subseteq \mbox{range}(f,{\bf x},I_{\cal A},I_{\cal E})$. 
\end{thm}

For instance, for a continuously $(n+1)$-differentiable $f$, the following $g$, obtained by a Taylor-Lagrange expansion and noting $x=(w,u)$, is an approximation function for $f$ 
\begin{multline}
g(x,\xi) = f(x^0)+\sum \limits_{i=1}^{n} \frac{(x-x^0)^i}{i !} D^i f (x^0)\\
+D^{n+1} f(\xi) \frac{(x-x^0)^{n+1}}{(n+1) !}
\label{innertaylor}
\end{multline}
\noindent where $D^\alpha f$ denotes the higher order partial derivative of $f$. For $n=0$, $g$ is the mean-value approximation.

\begin{pf}
We focus on the under-approximation. 
As $g$ is elementary, by Proposition 10.1 of \cite{gold2}, we have a continuous Skolem function 
$(w,u) \rightarrow \xi(w,u)$, i.e. a function such that for all $x=(w,u) \in \R^m$, $f(w,u)=g(w,u,\xi(x))$. 
Consider an under-approximation ${\cal I}_g$ of the robust range of $g$ with respect to $\xi$ and $x_{\cal A}$. It satisfies $\forall z \in {\cal I}_g, \ \forall w \in {\bf x}_{\cal A}, \ \forall \xi \in {\bf x}, \ \exists u \in {\bf x}_{\cal E}, \
z=g(w,u,\xi)$.
Let $z \in {\cal I}_g$ and $w \in {\bf x}_{\cal A}$ be fixed. 
As $g$ is elementary, we have a corresponding 
continuous Skolem function $\xi \rightarrow u(\xi)$, i.e. a function such that for all $\xi \in {\bf x}$, $z=g(w,u(\xi),\xi)$.
For this $z \in {\cal I}_g$ and $w \in {\bf x}_{\cal A}$, the continuous map $u\rightarrow u(\xi(w,u))$ 
defined from ${\bf x}_{\cal E}$ over itself, has a fixed point $u^\infty$, by Brouwer's theorem. It is such that
$z=f(w,u^\infty) = g(w,u^\infty, \xi(w,u^\infty))$. 
Hence $z$ is in the robust range of $f$ with respect to $x_{\cal A}$.
\end{pf}


\begin{example}
\label{ex:innertaylor2}
Consider function $f(x)=x^3+x^2+x+1$
on $[-\frac14,\frac14]$. The exact range is $[0.796875,1.328125]$. 
Let us approximate $f$ by a quadratic function, using an order 2 Taylor-Lagrange expansion. We compute
$f^{(1)}(x)  =  3x^2+2x+1$ and $f^{(2)}(x)  = 6x+2$.
By Theorem \ref{thm:generalAE}, the range of $f$ over $[-\frac14,\frac14]$ is under (resp. over) approximated by any under (resp. over) approximation of
the robust range with respect to $\xi$ of 
$$\begin{array}{rcl}
g(x,\xi)&=&f(x^0)+(x-x^0) f^{(1)}(x^0)+f^{(2)}(\xi) \frac{(x-x^0)^2}{2} \\
& = & 1+x+x^2 (3\xi+1)
\end{array}$$
\end{example}

In the general case, it may still be difficult to compute the under-approximated robust range of $g$. However, Theorem~\ref{thm:robustapprox} gives a simple way which is well suited in particular for quadratic Taylor-based approximations. 



\begin{thm}
\label{thm:robustapprox}
Let $g$ be an elementary function $g(w,u,\xi)=\alpha(w,u)+\beta(w,u,\xi)$ over $x=(w,u) \in {\bf x} \subseteq \I^m$ and $\xi \in {\bf x}$. Let
${\cal I}_{\alpha}$ be an under-approximation of the robust range of $\alpha$ with respect to $w$, i.e. $\mbox{range}(\alpha, {\bf x}, I_{\cal A}, I_{\cal E})$, and ${\cal O}_{\beta}$ an over-approximation of the range of $\beta$, i.e. $\mbox{range}(\beta,{\bf x}\times {\bf x},\emptyset, \{1,\ldots,2m\})$. 

The robust range  of $g$  with respect to $w \in {\bf x}_{\cal A}$ and $\xi
\in {\bf x}$, i.e. $\mbox{range}(g,{\bf x}\times {\bf x},I_{\cal A}\cup \{m+1,\ldots,2m\}, I_{\cal E})$, is under-approximated by
${\cal I}_{g} = [\underline{\cal I}_{\alpha}+\overline{\cal O}_{\beta},\overline{\cal I}_{\alpha}+\underline{\cal O}_{\beta}].$
\end{thm}

This is the case of Taylor expansions (\ref{innertaylor}), where $\alpha$ is the degree $n$ polynomial, and $\beta$ the degree $n+1$ remainder. 

\begin{pf}
${\cal I}_{\alpha}$ under-approximating $\mbox{range}(\alpha, {\bf x}, I_{\cal A}, I_{\cal E})$ means
$\forall a \in {\cal I}_{\alpha}, \ \forall w \in {\bf x}_{\cal A}, \ \exists u \in {\bf x}_{\cal E}, \ a= \alpha(w,u)$.
As $\alpha$ is elementary, 
we have a continuous Skolem function $(w,a)\rightarrow u(w,a)$.
%
%
Moreover, for all $z$ in ${\cal I}_{g}$, for all $b \in {\cal O}_{\beta}$, we have $z-b \in [\underline{\cal I}_{\alpha}+\overline{\cal O}_{\beta}-\overline{\cal O}_{\beta},\overline{\cal I}_{\alpha}+\underline{\cal O}_{\beta}-\underline{\cal O}_{\beta}]={\cal I}_{\alpha}$.

For given $z \in {\cal I}_{g}$,  $w \in {\bf x}_{\cal A}$ and $\xi \in {\bf x}$, consider the continuous function
$b \rightarrow \beta(w,u(w,z-b),\xi)$ over ${\cal O}_{\beta}$. By Brouwer fixed point theorem we have $b^\infty=\beta(w,u(w,z-b^\infty),\xi)$. Therefore, for any $z \in {\cal I}_{g}$, $\xi \in {\bf x}$, there exist $ a=z-b^\infty \in {\cal I}_{\alpha}$, $ b=b^\infty \in {\cal O}_{\beta}$ and $ x=(w,u(w,a)) \in {\bf x}$ such
that $z=a+b$. This implies that ${\cal I}_{g} $ is a robust under-approximation of $g(x,\xi)=\alpha(x)+\beta(x,\xi)$ with respect to $w$ and $\xi$.
\end{pf}

A direct consequence is a simple order 2 under-approxima\-ting Taylor method: 
\begin{corollary}
\label{corollary1}
Consider $f: \R^m \rightarrow \R$ a function in $C^2$. 
Let $\f^0$, $\bfm{\nabla}_{w}^0$ and  $\bfm{\nabla}_{u}^0$ be such that $ f(x^0) \subseteq \f^0  $, $ \left| \nabla_w f (x^0) \right| \subseteq  \bfm{\nabla}_{w}^0$ and $ \left| \nabla_u f (x^0) \right|  \subseteq  \bfm{\nabla}_{u}^0$ with $x^0=c(\x)$.

Then $\mbox{range}(f,\x,I_\mathcal{A},I_\mathcal{E})$  
is under-approximated by $[\underline{\cal I}_{\alpha}+\overline{\cal O}_{\beta},\overline{\cal I}_{\alpha}+\underline{\cal O}_{\beta}]$ where 
$
{\cal I}_{\alpha} = [ \overline{f^0}  - \langle \underline{\nabla}_u^0 ,  r({\x_\mathcal{E}}) \rangle + \langle \overline{\nabla}_w^0 , r({\x_\mathcal{A}}) \rangle ,$  $\underline{f^0} +  \langle \underline{\nabla}_u^0 , r({\x_\mathcal{E}}) \rangle - \langle \overline{\nabla}_w^0 ,  r({\x_\mathcal{A}}) \rangle ] 
$
\noindent and 
${\cal O}_{\beta}$ is any over-approximation of $\{\frac12 D^2 f(x) (r(\x))^2, x \in \x\}$. 
\end{corollary}


\begin{pf}
This is a direct application of Theorems~\ref{thm:generalAE} and \ref{thm:robustapprox} where $g$ is the 2nd order Taylor approximant, i.e. given by Equation (\ref{innertaylor}) for $n=1$, combined with Theorem~\ref{prop:robustAE} to compute the robust under-approximation of the order 1 approximation. 
We use a particular case of Theorem~\ref{prop:robustAE}: the under-approximation of an order 1 polynomial is almost trivial, we can compute it exactly if the computation is performed in real numbers with exact evaluation of $f$ and its gradient at a point. The expression we give here accounts for computation errors.
\end{pf}

\begin{example}
\label{ex:innertaylorbis}
We carry on with Example \ref{ex:innertaylor2}. The computation of the under approximation of the range of $1+x$ over $[-\frac14,\frac14]$ yields $[\frac34,\frac54]$. We also need an over approximation of the range of $x^2 (3\xi+1)$ for $x$ and $\xi$ in $[-\frac14,\frac14]$. Standard interval computation yields 
$[0,\frac{1}{16}][\frac14,\frac74]=[0,\frac{7}{64}]$. Overall, we deduce $[0.859375, 1.25] \subseteq \mbox{range}(f,\x)$. 
In comparison, the mean-value AE extension of Theorem \ref{prop:robustAE} would have given us the less precise under-approximation $[0.875, 1.125]$. 
\end{example}





\section{Refinements: preconditioning and quadrature formulae}
\label{sec:quadra}

The n-dimensional inner boxes that we compute with the techniques of Section~\ref{sec:vectorfunctions} can sometimes be small or empty, even when the projected inner-approximations on each component are tight. There are different reasons, for which we propose solutions in this section.

\subsection{Preconditioning for computing inner skewed boxes}
\label{sec:skewbox}
The first difficulty is when the image of the vector-valued function cannot be precisely approximated by a centered box.
\begin{example}
\label{ex_skewbox}
We consider 
$ f(x) = (  2 x_1^2  -  x_1 x_2 - 1 , x_1^2 + x_2^2 - 2 )^{\intercal} $
with $\x = [0.9,1.1]^2$. The under-approximated projections on the two components (respectively $[-0.38, 0.38]$ and $[-0.38, 0.38]$) are close to the over-approximated ranges  ($[-0.42, 0.42]^2$), but we only find empty inner boxes. 
\end{example}

This problem can be partly solved, as already described in \cite{lcss2020} by computing a skewed box as under-approximation, that is the image of a box by a linear map, 
instead of a  box. This can be achieved by combining  preconditioning to the mean-value theorem. Let $C \in \R^{n \times n}$ be a non-singular matrix. If  $\z$ is an interval vector such that $\z \subseteq \mbox{range}(C f,\x)$ , we can deduce a skewed box to be in the range of $f$, that is $\{C^{-1} z |  z \in \z \}   \subseteq \mbox{range}(f,\x)$.
A natural choice for $C$ is the inverse of the center of the interval Jacobian matrix  $C = (c(\bfm{\nabla}))^{-1}$. 

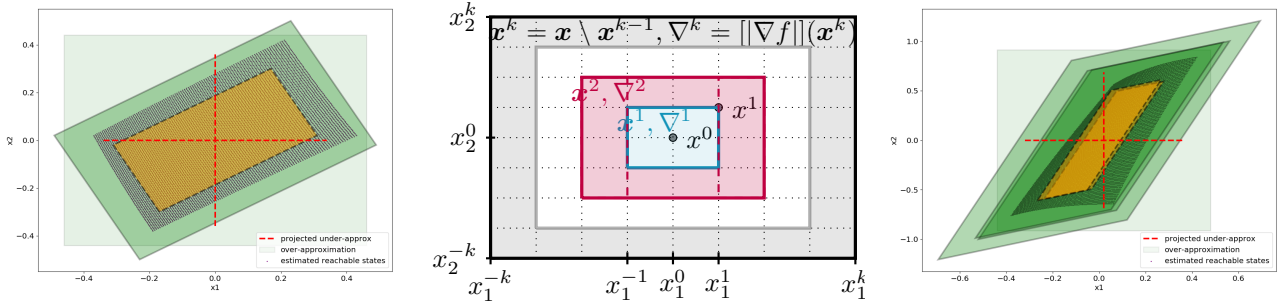
\begin{figure*}[htbp]
\begin{center}
\begin{subfigure}{0.3\textwidth}
\centering
\epsfig{file=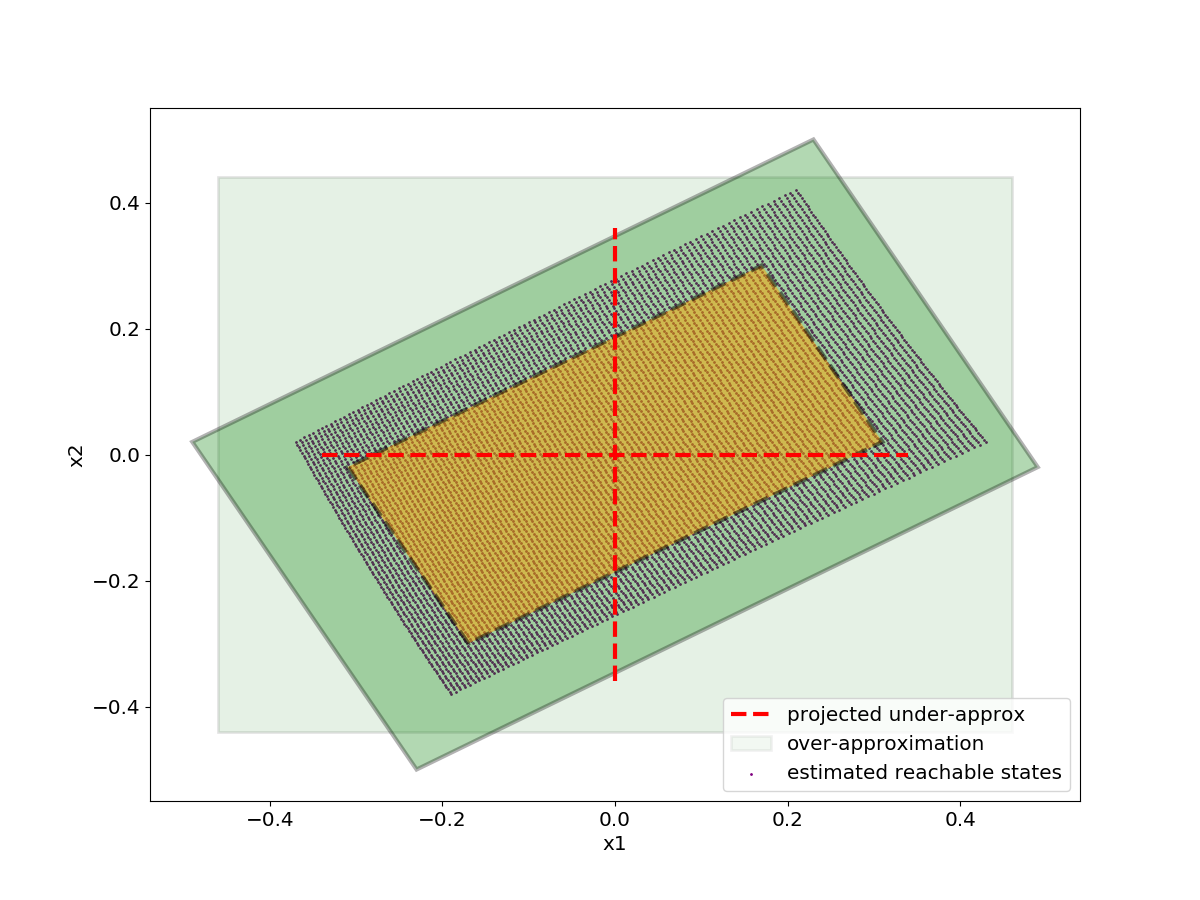,width=6cm}
\caption{Example~\ref{ex_skewbox_suite}: under- (dotted lines) and over-approximation (plain lines)}
\label{fig:ex_skewbox}
\end{subfigure}
~
\begin{subfigure}{0.3\textwidth}
\centering
\begin{tikzpicture}[xscale=0.6,yscale=0.4] 
    \tikzset{microphone/.style={black,circle,draw,fill=gray,scale=0.5,inner sep=2pt}};
	\node[microphone, label= right:$x^0$] at (4,4) (x0) {};
	\node[microphone, label= right:$x^1$] at (5,5) (x1) {};
    \coordinate (bottom_left) at (0,0);
    \coordinate (top_right) at (8,8);
    \draw [dotted, draw=black, fill=white] (bottom_left) grid  (top_right);
    
    \node[] at (4.0,7.5) {$\x^k=\x \setminus \x^{k-1}, \nabla^{k}=[\lvert \nabla f \rvert](\x^k)$};
    \draw[thick](0,-0.2)--(0,0.2) node[below=0.1]{$x_1^{-k}$};
    \draw[thick](3,-0.2)--(3,0.2) node[below=0.1]{$x_1^{-1}$};
    \draw[thick](4,-0.2)--(4,0.2) node[below=0.1]{$x_1^{0}$};
     \draw[thick](5,-0.2)--(5,0.2) node[below=0.1]{$x_1^{1}$};
    \draw[thick](8,-0.2)--(8,0.2) node[below=0.1]{$x_1^{k}$};
     \draw[thick](-0.1,0)--(0.1,0.) node[left=0.1]{$x_2^{-k}$};
    \draw[thick](-0.1,4)--(0.1,4) node[left=0.1]{$x_2^{0}$};
    \draw[thick](-0.1,8)--(0.1,8) node[left=0.1]{$x_2^{k}$};
    \draw[name path=A,very thick, draw=cyan!70!black,fill=cyan!70!black, fill opacity=0.1] (3,3) -- (3,5) -- (5,5) -- (5,3) -- (3,3);
    \node[color=cyan!70!black] at (3.6,4.5) {$\x^1, \nabla^{1}$};
     \draw[name path=B,very thick, draw=purple] (2,2) -- (2,6) -- (6,6) -- (6,2) -- (2,2);
     \draw[dashed,thick, draw=purple] (2,2) -- (3,2) -- (3,6) -- (2,6) -- (2,2);
      \draw[dashed,thick, draw=purple] (5,2) -- (6,2) -- (6,6) -- (5,6) -- (5,2);
      
     \tikzfillbetween[of=A and B]{purple, opacity=0.2};
  	 \draw [name path=D,very thick, draw=black] (0,0) -- (0,8) -- (8,8) -- (8,0) -- (0,0);
     \draw[name path=C,very thick, draw=white!70!black] (1,1) -- (1,7) -- (7,7) -- (7,1) -- (1,1);
      \tikzfillbetween[of=D and C]{black, opacity=0.1};
    \node[color=purple] at (2.6,5.5) {$\x^2, \nabla^{2}$};
    \end{tikzpicture}
\caption{Partitioning the input domain \label{fig:partition}}
\end{subfigure}
~
\begin{subfigure}{0.3\textwidth}
\centering
\epsfig{file=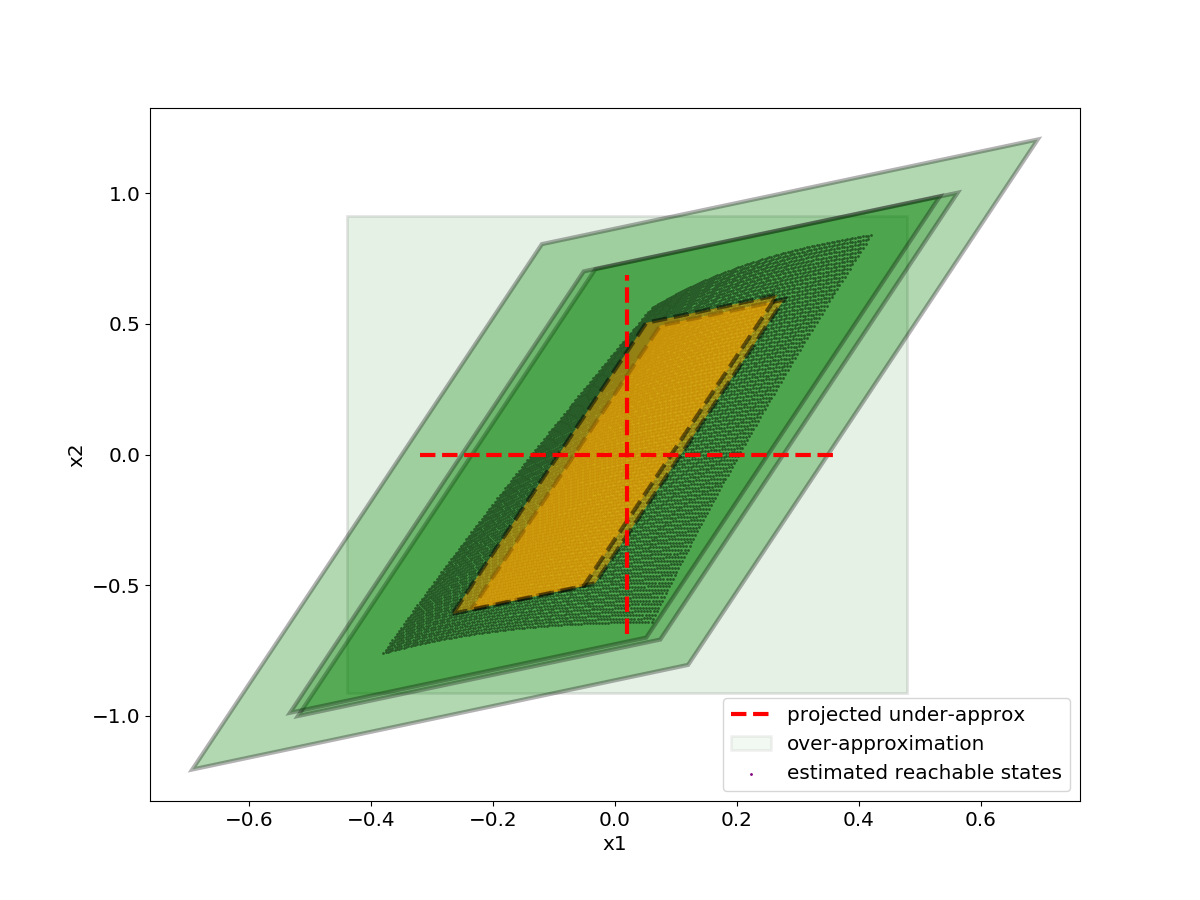,,width=6cm}
\caption{Example~\ref{ex:partition}: approximations for quadrature and order 2 extensions}
\label{fig:expartition}
\end{subfigure}
\caption{Illustrations for Sections~\ref{sec:vectorfunctions} and \ref{sec:quadra}}
\label{fig:car}
\end{center}
\end{figure*}

\begin{example}
\label{ex_skewbox_suite}
On Example~\ref{ex_skewbox}, using this preconditioning and $pi : (1 \rightarrow 1, 2 \rightarrow 2)$, 
we obtain for $f_2$ as a function of $f_1$ the yellow under-approximating parallelotope of Figure~\ref{fig:ex_skewbox}.  
We estimate the image $\mbox{range}(f,\x)$ by sampling points in the input domain. This sampling-based estimation is represented as the dark dots-filled region. The  green parallelotope and box are the over-approximations with and without preconditioning. 
\end{example}

%
%
%
%
%

\subsection{Quadrature formulae for the mean-value extension}
\label{sec:quadrature}
The mean-value interval extension can yield a rough approximation. This is especially the case when the variation of the gradient is important over the input range, the extreme case for under-approximation being when this variation contains zero: the under-approximation is empty or reduced to a point. Using simple quadrature formulae partially solves this problem. 

Let $f: \R^m \rightarrow \R$.
 We partition each dimension $j=[1 \ldots m]$ of the $m$-dimensional input box $\x=\x_1 \times \ldots \times \x_m$ in $2k$ sub-intervals and define, for all $j=[1 \ldots m]$,  $x_j^{-k} \leq x_j^{-(k-1)} \leq \ldots \leq x_j^0 \leq \ldots \leq x_j^k$, with
$x_j^{-k}=\underline{\x_j}$, $x_j^0=\mbox{mid}({\x_j})$, $x_j^{k}=\overline{\x_j}$. We note $dx^i = x^{i} - x^{i-1}$ the vector-valued deviation. 

Let us refine the  mean-value AE extensions using such a partition. 
The first natural idea is to compute an under-approximation for each sub-box obtained as product of sub-intervals in each dimension. 
But in general, the under-approximating boxes will be non-contiguous, and their convex union is in general not an under-approximation of $ \mbox{range}(f,\x)$. 
Moreover, this approach would not scale well. 

We now propose a scheme that avoids these unions, and remains linear  in $k$ with respect to the non-partitioned case. 
We note $\x^1=[x_1^{-1},x_1^{1}] \times [x_2^{-1},x_2^{1}] \times \ldots \times  [x_m^{-1},x_m^{1}]  $, and for all $i$ between $2$ and $k$, $\x^i=[x_1^{-i},x_1^{i}] \times \ldots \times [x_m^{-i},x_m^{i}] \setminus \mathring{\x}^{i-1}$, where $\setminus$ denotes the set difference and $\mathring{\x}$ the interior of $\x$. This partition is represented in Figure~\ref{fig:partition} for a two-dimensional input space. In practice,  each "square ring" $\x^i$ will be decomposed in $2n$ sub-boxes for the Jacobian evaluation.

By the mean-value theorem,  $\forall x \in [x^{-1},x^1]$, $\exists \xi^1 \in [x^{-1},x^1],$ 
$f(x) = f(x^0) + \langle \nabla f(\xi^1) ,x-x^0 \rangle $. Suppose we can compute  
 $\f^0 \supseteq f(x^0)$ and  $\bfm{\nabla}^i$ for $i$ in $[1,k]$ such that $\{ \lvert \nabla f(x) \rvert, x \in \x^i \} \subseteq \bfm{\nabla}^i$. 
We have 
$\mbox{range}(f,\x^1) \subseteq  \f^0 +  \langle \overline{\nabla}^1 , dx^1 \rangle [-1,1]$
and $[\overline{f^0} - \langle \underline{\nabla}^1 , dx^1 \rangle , \underline{f^0} +  \langle \underline{\nabla^1} , dx^1 \rangle ] \subseteq \mbox{range}(f,[x^{-1},x^{1}])$.
Let us now take $x \in \x^2$. 
We can iterate the mean-value theorem on the adjacent input subdivision and write that for all $x \in \x^2$, there exists $x^1 \in  \x^1 \cap \x^2$ (that is on the border between $\x^1$ and $\x^2$), there  exists $\xi^2 \in \x^2$ such that $f(x) = f(x^1) + \langle  \nabla f(\xi^2) , x-x^1 \rangle $
 and $\lvert x_1-x^1_1 \rvert \leq dx_1^2$ and $\lvert x_2-x^1_2 \rvert \leq dx_2^2$. (take for example for $x^1$ the intersection of the line from $x^0$ to $x$ with the border between $\x^1$ and $\x^2$). 
%
We have 
$ \mbox{range}(f,\x^1 \cup \x^2) \subseteq \f^0 +  \langle \overline{\nabla}^1 , dx^1 \rangle [-1,1] +  \langle \overline{\nabla}^2 , dx^2 \rangle [-1,1].$
There also exists $(x,x^1) \in \x^2 \times \x^1$ such that  $\lvert x_1-x^1_1 \rvert = dx^2_1$ and $\lvert x_2-x^1_2 \rvert = dx^2_2$ (take the corners of the boxes $\x^1$ and $\x^2$), so that we also have
$ [\overline{f^0} - \langle \underline{\nabla}^1 , dx^1 \rangle - \langle \underline{\nabla}^2 , dx^2 \rangle, \underline{f^0} + \langle \underline{\nabla}^1 , dx^1 \rangle + \langle \underline{\nabla}^2 , dx^2 \rangle ] \subseteq \mbox{range}(f,\x^1 \cup \x^2). $

This generalizes to the $k$ subdivisions:
\begin{align}
 \mbox{range}(f,\x) \subseteq  \f^0 +  \sum_{i=1}^k  \langle \overline{\nabla}^i , dx^i \rangle [-1,1]
 \label{eq:subdiv1} \\
 [\overline{f^0} - \sum_{i=1}^k \langle \underline{\nabla}^i , dx \rangle , \underline{f^0} + \sum_{i=1}^k  \langle \underline{\nabla}^i , dx \rangle ] \subseteq \mbox{range}(f,\x) \label{eq:subdiv2}
\end{align}
The same idea applies to the estimation of robust ranges. 

Naturally, other schemes can be proposed, relying on the idea that this technique can be seen as using a quadrature formula for integrating the Jacobian of a function.


\begin{example}
\label{ex:partition} 
We consider 
$ f(x) = (  2 x_1^2 + 2 x_2^2  - 2  x_1 x_2 - 2  , x_1^3 - x_2^3 + 4 x_1 x_2 - 3 )^{\intercal} $
with $\x = [0.9,1.1]^2$. 
The results are represented in Figure~\ref{fig:expartition}. The sampling-based estimation of the image is the dark dots-filled region. We choose $pi : (1 \rightarrow 1, 2 \rightarrow 2)$. 
Using the preconditioned mean-value extension without partitioning, the over-approximation is the largest green parallelotope and the under-approximation for the joint range is empty.
The quadrature formula for the mean-value extension with $k=10$ partitions on one hand, and the order 2 extension of Corollary~\ref{corollary1} on the other hand, yield two very similar under-approximating yellow parallelotopes. They also yield two very similar green over-approximating parallelotopes. Both approaches actually have comparable precision on the different examples tested. The light green box is the order 2 over-approximation without preconditioning. 
\begin{remark}
One could be tempted to use more partitions to improve the quality of the approximation. However, the computations yield approximations that are centered at 
$f(x^0)$.  We can observe that the under-approximating skewed box is actually already very close to being the largest skewed box entirely included in the image, given a fixed skewing and a center at $f(x^0)$.
The order 2 estimation  allows for slight decentering, but it would also be possible to use less basic quadrature formulae for that purpose. 
Quadrature can also  be combined with  order 2 extensions.
\end{remark}

\end{example}
Quadrature can also be combined with some classical partition of the inputs, or with different center choices. However, a possibly disjoint union of approximations (corresponding to a classical partition of inputs) is inconvenient, so we would recommend this use only as a property-driven refinement. Finally, as a sound under-approximation is still obtained by considering a sub-region of the input set, refinements can be obtained by detecting and removing sub-regions where Jacobian coefficients are either very sensitive to the inputs or close to zero.


\subsection{Bounding the Jacobian matrix}
The approach relies on being able to compute over-approximations of $\nabla f(x)$ over some sub-sets of input box $\x$, namely  $\bfm{\nabla}^i$ for $i$ in $[1,k]$ such that $\{ \lvert \nabla f(x) \rvert, x \in \x^i \} \subseteq \bfm{\nabla}^i$. Automatic differentiation allows to compute the derivatives, but need to be combined with set-membership methods to handle uncertainties. We have found the combination of automatic differentiation with an evaluation in affine arithmetic to provide a good trade-off between efficiency and precision. All the experiments presented in this work were performed with an implementation relying on this combination. Affine forms provide an interesting combination of parameterization and set-based estimation:  a parametric approximate form for $\nabla f(x)$ valid on all box $\x$ is computed, that can be instantiated on $\x^i$ to yield tight over-approximations  $\nabla^i$, without the need for performing several evaluations of the differentiation.
The full description is out of scope,  we give below a flavor of the use of affine arithmetic for Jacobian estimation on a simple example. 
\begin{example}
Consider in Example~\ref{ex:partition} the derivative of $f_1(x_1,x_2)=2 x_1^2 + 2 x_2^2  - 2  x_1 x_2 - 2$ with respect to $x_1$, with $(x_1,x_2) \in [0.9,1.1]^2$. This derivative is $\nabla_1 f_1 (x_1,x_2)= 4 x_1 - 2 x_2$. Evaluation with affine arithmetic first consists in creating a centered form with a fresh noise symbol for each input: $\hat x_1 = 1. + 0.1 \varepsilon_1$ and  $\hat x_2 = 1. + 0.1 \varepsilon_2$, with $(\varepsilon_1, \varepsilon_2)\in [-1,1]^2$. The  gradient evaluated on these affine forms is 
$\hat \nabla_1 \hat f_1 (x)= 2 + 0.4 \varepsilon_1 - 0.2  \varepsilon_2.$
Here the abstraction is exact. In the general case, affine arithmetic will compute an approximate affine form and bound the approximation error in a new noise term. 
Let us now use this affine form to over-approximate $ \nabla_1 f_1(x)$ over all $\x$. This amounts to computing the interval bounds when $\varepsilon_1$ and $\varepsilon_2$ range  in $[-1,1]$, for which we get 
$\bfm{\nabla}_{1,1} = [1.4,2.6]$. Now we use the same affine form to over-approximate $\nabla_1 f_1(x)$ over $\x^1$. This amounts to take  $\varepsilon_1$ and $\varepsilon_2$ both ranging  in $[-\frac1k,\frac1k]$. Let us consider $k=10$ subdivisions, we obtain 
$\bfm{\nabla}^1_{1,1} = 2+0.4[-0.1,0.1]-0.2[-0.1,0.1]=[1.94,2.06]$. The process can be iterated to compute  $\bfm{\nabla}^i_{1,1}$ for $i \in [2 \ldots k]$, expressing the membership to a subset $\x^i$ of $\x$ as constraints on $\varepsilon_1$ and $\varepsilon_2$, used to instantiate to $\x^i$ the estimation of $\nabla_1 f_1(x)$. For instance, in order to compute $\bfm{\nabla}^2_{1,1}$, we decompose the computation on the 4 rectangles that define $\x_2$ (see Figure~\ref{fig:partition}):
$(\varepsilon_1,\varepsilon_2) \in ([\frac1k, \frac2k] \times [-\frac2k,\frac2k]) \cup ([-\frac2k, -\frac1k] \times [-\frac2k,\frac2k]) \cup ( [-\frac1k, \frac1k] \times [-\frac2k, -\frac1k])  \cup ( [-\frac1k, \frac1k] \times [\frac1k, \frac2k]) $.
\end{example}



\section{Application to the reachability of discrete-time systems}
\label{sec:discrete_systems}

We consider discrete-time non-linear dynamical systems with inputs of the form
\begin{equation}
\begin{cases}
 z^{k+1} = f(z^k, u^k) \\
 z^0 \in \z^0 
 \end{cases}
 \label{eq:discretesys}
 \end{equation}
where $f: \R^m \rightarrow \R^n$  is a vector-valued non-linear function with 
$m \geq n$,  $z \in \R^n$ the vector of state variables, $u \in {\bfm{u}} \subseteq \R^{m-n}$ 
the input signal, and $\z^0$ the initial set. 

Given an initial set $\z^0$, we want to compute the bounded time reachable set of the dynamical system, i.e, the set of states visited by the dynamical system up to a fixed
time horizon $K \in \N$. The reachable set can be obtained
as the solution of the recursion  $\z^{k+1} = \{f(z^k,u^k) | z^k \in \z^k, u^k \in {\bfm{u}} \}$,
for $k \in [ 0,K]$. The computation of the
reachable set can be seen as a series of images of sets by vector-valued function $f$. We thus can use the results of Sections~\ref{section1} to  \ref{sec:quadra} to compute approximations of these reachable sets.

For conciseness, we consider  systems without disturbances 
and compute maximal (or classical) reachable sets. The algorithms can be straightforwardly extended to  robust reach set of systems with disturbances, basically replacing ranges by robust ranges. 
This allows us to use the lighter notations ${\cal I}(f,\x,\pi)$ and ${\cal O}(f,\x,\pi)$ to note the under and over-approximating sets introduced in Definition~\ref{def1}. 


\subsubsection{Method 1}
\label{method1}
the first method consists in iteratively using function image, independently for under and 
over-approximation, taking as input the previously computed approximation of the image.
We compute under and over-approximations $I^k$ and $O^k$ of the reachable set $\z^{k}$ 
by  
\begin{equation}
\begin{cases} I^0=\z^0, \; O^0=z^0 \\
I^{k+1}={\cal I}(f,I^k,\pi), \; O^{k+1}={\cal O}(f,O^k,\pi)
\end{cases}
\end{equation}
Indeed, at each step $k$, we have $I^{k+1} \subseteq \mbox{range}(f,I^k) \subseteq \mbox{range}(f,\z^k) = \z^{k+1} \subseteq \mbox{range}(f,O^k) \subseteq O^{k+1}$.

With this approach, at each step $k$, under and over-approximations $I^{k}$ and $O^{k}$ of the  joint range are used as input for the next step. It is thus  particularly important to compute tight under and over-approximations of this joint range. In particular, using the preconditioning of Section~\ref{sec:skewbox} will often be crucial, both for under and over approximation. 
This yields Algorithm~\ref{alg:discrete1}.
\begin{algorithm}
\caption{Iterated discrete-time reachability}
\label{alg:discrete1}
\begin{algorithmic}
\REQUIRE $f : \R^n \rightarrow \R^n$, $\z^0 \subseteq \I^n$ initial state, $K \in \N^+$,  an over-approximating extension $[\nabla f]$ (see Section~\ref{sec:quadra})
\ENSURE  $I^k$ and $O^k$: under and over-approximations  of the reachable set $\mbox{range}(f^k,\z^0)$ for $k \in [1,K]$
\STATE $I^0 \coloneqq \z^0, O^0 \coloneqq \z^0$; choose $\pi: [1 \ldots n] \mapsto [1 \ldots n]$
\FOR{$k$ from 0 to $K-1$} 
\STATE $\bfm{\nabla}_{I}^k \coloneqq | [\nabla f] (I^{k}) | $, $\bfm{\nabla}_{O}^k \coloneqq | [\nabla f] (O^{k}) | $ 
\STATE $A_{I}^k \coloneqq c(\bfm{\nabla}_{I}^k)$, $A_{O}^k \coloneqq c(\bfm{\nabla}_{O}^k)$ (supposed non-singular, otherwise taken to identity matrix)
\STATE $C_{I}^k \coloneqq (A_{I}^k)^{-1}$, $C_{O}^k \coloneqq (A_{O}^k)^{-1}$ 
\STATE $ \z_{I}^{k+1} \coloneqq {\cal I}(C_I^k f,I^k,\pi)$, $ \z_{O}^{k+1} \coloneqq {\cal O}(C_O^k f,O^k,\pi)$
\IF{$\z_{I}^k = \emptyset$}
\STATE return
\ENDIF
\STATE $I^{k+1} \coloneqq A_{I}^k  \z_{I}^{k+1} $, $O^{k+1} \coloneqq A_{O}^k  \z_{O}^{k+1} $
\ENDFOR
\end{algorithmic}
\end{algorithm}
At each step $k=0 \ldots K-1$, $\z_I^{k+1}$ is an interval vector such that, if it is non empty, $I^{k+1} = A_I^k \z_I^{k+1} \subseteq \mbox{range}( f, I^k)  \subseteq\mbox{range}(f^{k+1},\z^0)$. The over-approximation is computed similarly, and is fully decoupled.

\subsubsection{Method 2}
\label{methode2}
the second method consists in computing the sensitivity to initial state by approximating the gradient of the iterated function. At each step $k$, we compute the under and over-approximation of $\mbox{range}(f^k,\z^0)$, i.e. the loop body $f$ iterated $k$ times, starting from the initial state $\z^0$. 
This yields the schematic Algorithm~\ref{alg:discrete2}, with same inputs and hypotheses as in Algorithm~\ref{alg:discrete1}.
\begin{algorithm}
\caption{Discrete-time reachability computed on $f^k$}
\label{alg:discrete2}
\begin{algorithmic}
\FOR{$k$ from 0 to $K-1$} 
\STATE $ I^{k+1} \coloneqq {\cal I}(f^{k+1},\z^0,\pi)$, $ O^{k+1}  \coloneqq {\cal O}(f^{k+1},\z^O,\pi)$
\ENDFOR
\end{algorithmic}
\end{algorithm}
Here, at each step $k$, the under- and over-approximation are both obtained from an over-approximation of $f^{k+1}$ evaluated at the center of $\z^0$ and an over-approximation of the gradient of $f^{k+1}$ over $\z^0$: at each step, the gradient can be obtained by differentiating the gradient from the previous step.  Of course, this can also be combined with preconditioning. 

\subsubsection{Discussion}
While relying on the same techniques for range estimation, Algorithm~\ref{alg:discrete1} and Algorithm~\ref{alg:discrete2} are  different in spirit: Algorithm~\ref{alg:discrete2} relies only on the propagation of over-approximations to deduce under-approximations. In particular, the under-approximation may be empty at some step $k$, and become non-empty again at further steps (a similar remark was made in the context of continuous systems in \cite{hscc2017}). In comparison, Algorithm~\ref{alg:discrete1} needs at each step an under-approximating box or skew box that is non-empty for all components. 
On the other hand, Algorithm~\ref{alg:discrete2} is more costly as it requires a differentiation of the iterated function. 

\section{Implementation and examples}

\begin{figure}[htbp]
\centerline{\epsfig{file=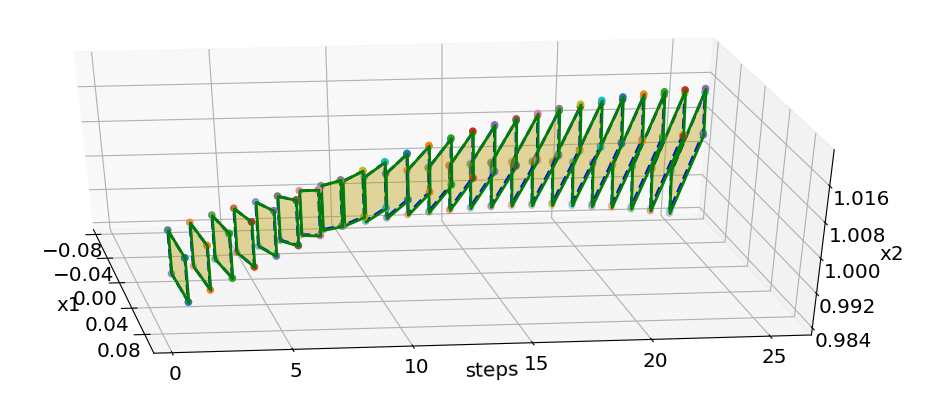,clip=,width=9.5cm, height=4.5cm}}
\caption{\label{fig:testmodel4} Skewed box under and over-approximations  for 25 steps (Algorithm~\ref{alg:discrete1}) for the test model}
\end{figure}

We now present  results on small systems from the literature. The approach is implemented as part of the RINO C++ prototype, available from \url{https://github.com/cosynus-lix/RINO}. The prototype allows to experiment the function range estimation, but its actual target is 
discrete and continuous-time  reachability, combining for continuous-time the techniques presented here with Taylor model methods~\cite{hscc19,lcss2020}. The prototype uses the fadbad++ (\url{http://www.fadbad.com/fadbad.html}) automatic differentiation library and the aaflib (\url{http://aaflib.sourceforge.net/}) affine arithmetic library. 
The timings are given on a Macbook Pro 2.6GHz Intel Core i7 and 32Gb of RAM.
\paragraph*{Test Model}
We consider the test model~\cite{Dreossi_2016}:
\[
\label{eq:testmodel}
\begin{split}
x_1^{k+1}& = x_1^k + (0.5 (x_1^k)^2 - 0.5 (x_2^k)^2) \Delta \\
x_2^{k+1}& = x_2^k + 2 x_1^k x_2^k  \Delta
\end{split}
\]
with as initial set a box $x_1 \in [0.05, 0.1]$ and $x_2 \in [0.99, 1.00]$, and $\Delta = 0.01$.
Figure \ref{fig:testmodel4} shows the under and over-approximated reachable sets (respectively the filled yellow region and green parallelotope) over time up to 25 steps with Algorithm~\ref{alg:discrete1}. They are obtained in 0.02 seconds. We can observe that the under and over-approximations are very close one to another, confirming the accuracy of the results.  


\paragraph*{SIR Epidemic Model}
We now consider the SIR epidemic model with the parameters of \cite{Dreossi_2016}.  
\[
\begin{split}
x_1^{k+1} &= x_1^k - \beta   x_1^k x_2^k \Delta \\
x_2^{k+1} &= x_2^k + (\beta x_1^k  x_2^k - \gamma x_2^k)  \Delta \\
x_3^{k+1} &= x_3^k + \gamma x_2^k \Delta 
\end{split}
\]
We compute the reachable set up to 60 steps from the initial box $(x_1,x_2,x_3) \in [0.79, 0.80] \times  [0.19, 0.20] \times [0,0.1]$. The  parameter values are $\beta = 0.34$, $\gamma = 0.05$, and $\Delta = 0.5$.

\begin{figure}
\centerline{\epsfig{file=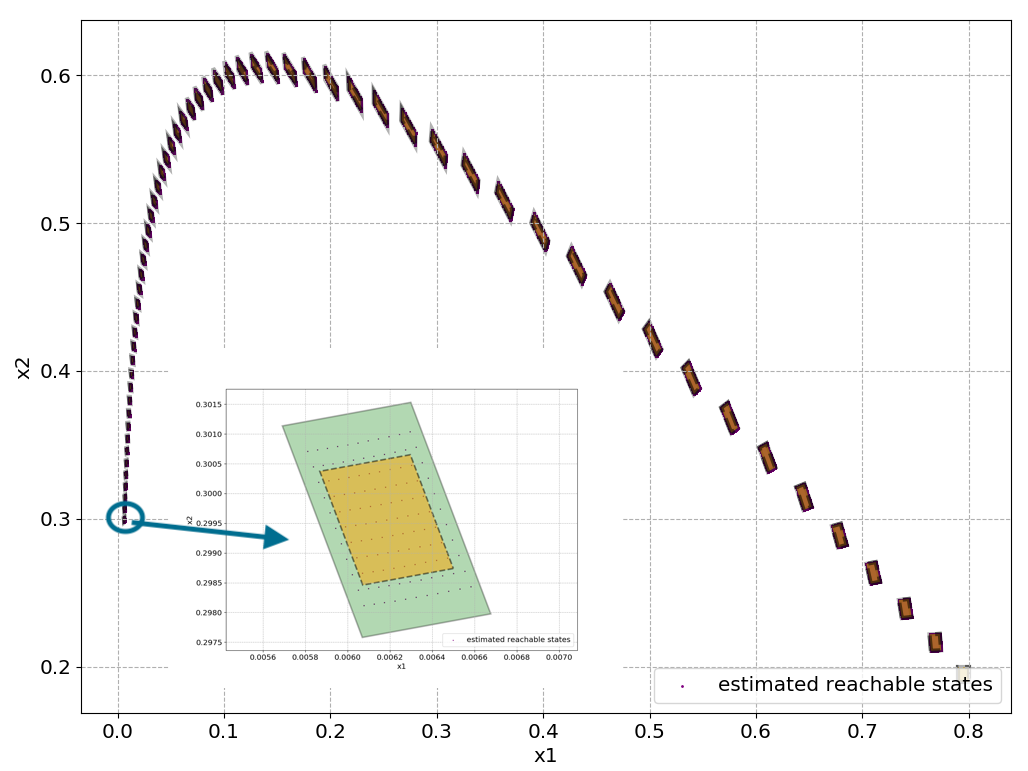,clip=,width=9cm, height=4cm}}
\caption{\label{fig:SIR1} Skewed box under and over-approximations  for 60 steps (Algorithm~\ref{alg:discrete1}) of the SIR epidemic model}
\end{figure}

The reachable sets computed in 0.05 seconds with  Algorithm~\ref{alg:discrete1} up to 60 steps are represented  for $(x_1,x_2)$ in Figure~\ref{fig:SIR1}. 
We can note in particular from the zoomed reachable set in the Figure, which corresponds to last step (60), that the under-approximation (in yellow) is still of good quality (the purple dots correspond to sample executions).

\begin{figure}
\begin{subfigure}{.48\linewidth}
\hspace{-.1cm}\epsfig{file=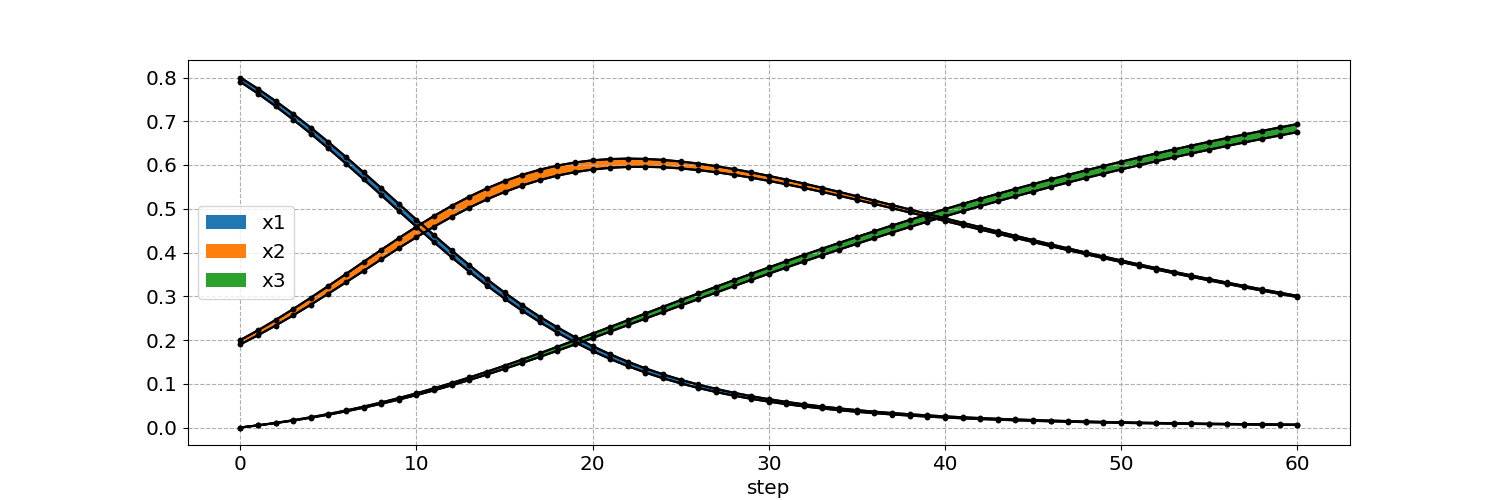,clip=,width=5cm,height=3.5cm} 
\caption{SIR epidemic model, 60 steps}
\label{fig:SIR4}
\end{subfigure}
\begin{subfigure}{.45\linewidth}
\epsfig{file=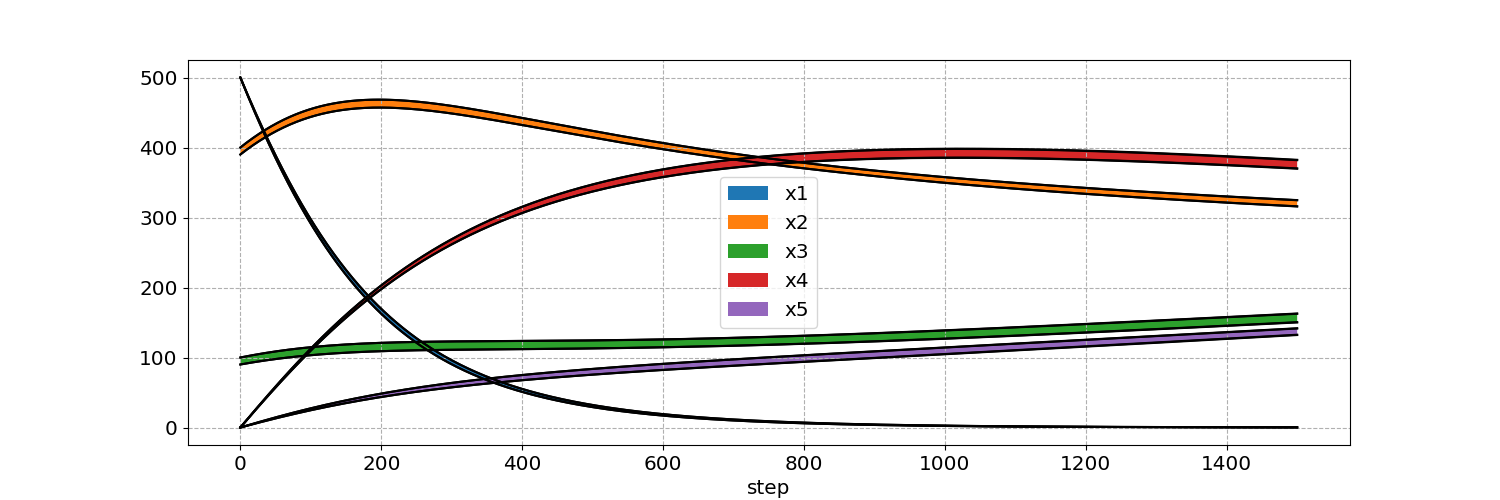,clip=,width=5cm,height=3.5cm}
\caption{Bees model, 1500 steps}
\label{fig:Bees3}
\end{subfigure}
\caption{Projected under and over-approximations, Algorithm~\ref{alg:discrete2}}
\label{fig2}
\end{figure}

However, only Algorithm~\ref{alg:discrete2} is able to compute non-empty approximations when 
taking as initial condition  $x_3 = 0$ which is of empty interior, instead of $x_3  \in [0,0.1]$. We obtain in 0.05 seconds the very tight approximations of the projections of the components represented Figure\ref{fig:SIR4}. 

\paragraph*{Honeybees Site Choice Model}
We consider the reachable sets up to 1500 steps of the model studied in \cite{Dreossi_2016}:
\[
\begin{split}
x_1^{k+1} &= x_1^k - (\beta_1 x_1^k  x_2^k + \beta_2 x_1^k x_3^k) \Delta \\
x_2^{k+1} &= x_2^k + (\beta_1 x_1^k  x_2^k  - \gamma  x_2^k + \delta \beta_1  x_2^k x_4^k + \alpha \beta_1 x_2^k x_5^k)  \Delta \\
x_3^{k+1} &= x_3^k + (\beta_2 x_1^k x_3^k - \gamma x_3^k + \delta \beta_2 x_3^k x_5^k + \alpha \beta_2 x_3^k x_4^k)  \Delta \\
x_4^{k+1} &= x_4^k + (\gamma x_2^k - \delta \beta_1  x_2^k x_4^k - \alpha \beta_2 x_3^k x_4^k)  \Delta \\
x_5^{k+1} &= x_5^k + (\gamma x_3^k - \delta \beta_2 x_3^k x_5^k -  \alpha \beta_1 x_2^k x_5^k)  \Delta \\
\end{split}
\]
with as initial set the box $x_1 = 500$, $x_2 \in [390, 400]$, $x_3 \in [90, 100]$, $x_4 = x_5 = 0$ and the parameter values $\beta_1 = \beta_2 = 0.001$, $\gamma = 0.3$, $\delta = 0.5$, $\alpha = 0.7$, and $\Delta = 0.01$.

Algorithm~\ref{alg:discrete1} runs very fast, taking only 1.7 seconds for the 1500 steps. But the under-approximation is very soon empty and  the over-approximation tends to strongly widen after 800 steps. 
%
In comparison, Algorithm~\ref{alg:discrete2} takes 57 seconds to complete the 1500 reachability steps. 
But the projected under-approximations are very tight, close to the over-approximations, as can be seen on  Figure~\ref{fig:Bees3} which represents under and over-approximations for all components as functions of steps. 
These results should also be compared to the much wider over-approximation of \cite{Dreossi_2016} (Figure 7, where time is number of steps divided by 100), obtained in 81 seconds.




\section{Conclusion and future work}
We focused on new AE under-approximating extensions and their accurate practical evaluation for non-linear vector-valued functions, and exemplified their interest for the reachability of discrete-time systems. These techniques can  also be used for the reachability analysis of continuous-time systems, improving for instance over ~\cite{hscc19,lcss2020}.


\bibliography{IEEEabrv,emsoft2020}

\end{document}